\documentstyle[prb,aps,epsf,multicol]{revtex}

\tighten
\begin{document}
\draft

\title{Zero-point fluctuations in the ground state of a mesoscopic
normal ring}

\author{Pascal Cedraschi and Markus B\"uttiker}
\address{D\'epartement de Physique Th\'eorique,
Universit\'e de Gen\`eve,\\
24, quai Ernest Ansermet, CH-1211 Geneva 4, Switzerland}

\date{\today}

\maketitle

\begin{abstract}
We investigate the persistent current of a ring with an in-line
quantum dot capacitively coupled to an external circuit.  Of special
interest is the magnitude of the persistent current as a function of
the external impedance in the zero temperature limit when the only
fluctuations in the external circuit are zero-point fluctuations.
These are time-dependent fluctuations which polarize the ring-dot
structure and we discuss in detail the contribution of displacement
currents to the persistent current.  We have earlier discussed an
exact solution for the persistent current and its fluctuations based
on a Bethe ansatz. In this work, we emphasize a physically more
intuitive approach using a Langevin description of the external
circuit.  This approach is limited to weak coupling between the ring
and the external circuit. We show that the zero temperature persistent
current obtained in this approach is consistent with the persistent
current calculated from a Bethe ansatz solution.  In the absence of
coupling our system is a two level system consisting of the ground
state and the first excited state.  In the presence of coupling we
investigate the projection of the actual state on the ground state and
the first exited state of the decoupled ring. With each of these
projections we can associate a phase diffusion time. In the zero
temperature limit we find that the phase diffusion time of the excited
state projection saturates, whereas the phase diffusion time 
of the ground state projection diverges.

\end{abstract}

\pacs{PACS numbers: 73.23.Ra, 73.23.Hk, 71.27.+a}

\begin{multicols}{2}
\narrowtext

\section{Introduction}
\begin{figure}[ht]
\centerline{\epsfysize=6cm\epsfbox{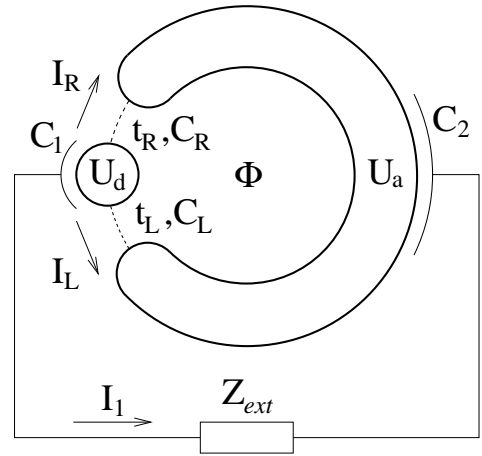}}
\vspace*{0.3cm}
\caption{\label{deph:system}Ring with an in-line dot subject to a flux
$\Phi$ and capacitively coupled to an external impedance $Z_{ext}$.}
\end{figure}

An interesting aspect of quantum systems is the fact that even at zero
temperature there are fluctuations which manifest themselves in
obser\-vables which do not commute with the Hamiltonian of the system.
In mesoscopic systems phase coherence plays an essential role, and it
is thus important to ask to what extent coherence is affected by
zero-point fluctuations. To investigate this question we consider the
{\em ground state\/} of a normal mesoscopic ring threaded by an
Aharonov-Bohm flux and capacitively coupled to an external
circuit. Such a ring exhibits a persistent current which is a direct
consequence of phase coherent electron motion over distances large
compared to the ring circumference.  At zero temperature such a ring
interacts with an external circuit (see Fig.~\ref{deph:system}) only
due to zero-point fluctuations. More precisely, the external circuit
can, through the generation of zero-point voltage fluctuations, induce
polarization fluctuations in the ring which in turn affect the
magnitude of the persistent current.  The source of the voltage
fluctuations in the external circuit are the resistive elements. We
can thus ask: How does the persistent current of the ring depend on
the resistive properties of the external circuit? For the system shown
in Fig.~\ref{deph:system}, Ref.~\onlinecite{cedraschi:prl} provided an
answer by mapping a simple model of a ring with a quantum dot and
external circuit on the anisotropic Kondo model and using the known
Bethe ansatz solutions of this problem.\cite{thesis} The purpose of this 
work is
to consider the same model and to provide a discussion which is
physically more transparent. The discussion given below is, however,
limited to the case of weak coupling between the ring and the external
circuit.

Investigations of the coherence properties of the ground state of
mesoscopic structures are rare and concern mainly superconductors.
Hekking and Glazman\cite{hekking:loop} investigate a thin
superconducting loop in an electromagnetic environment; Oshikawa and
Zagoskin\cite{zagoskin:grainwire} consider a superconducting grain
coupled to a normal wire. In contrast, investigations on dephasing in
systems driven out of the ground state are numerous and there exists a
considerable literature. In particular, zero-point fluctuations have
been of interest, following recent experimental work by Mohanty,
Jariwala, and Webb\cite{mohanty:decoh} on weak localization in
metallic diffusive wires. These experiments probe a transport state
and not the ground state of the system.  In the experiments of Mohanty
and Webb it is the weak-localization effect in the linear response
conductance of the system which is of interest. Subsequently to these
experiments the role of zero-point fluctuations in weak localization
effects has been very much debated\cite{golubev:decoh,aleiner:decoh,%
aleiner:decoh:comment,golubev:decoh:reply} and a number of works have
appeared which suggest that the experiments can possibly be better
explained without invoking zero-point fluctuations.\cite{kravtsov:pc}
Weak-localization is a quantum effect which survives ensemble
averaging and thus the dephasing rates which count are specific to the
fact that we deal with time-reversed trajectories and that an ensemble
average has to be performed. The absence of an effect of zero-point
fluctuations found in Refs.~\onlinecite{aleiner:decoh,%
aleiner:decoh:comment} might give rise to the mistaken impression that
zero-point fluctuations are quite generally without any effect on the
coherence properties of a system.  The example presented here shows
that zero-point fluctuations are clearly important for the coherence
properties of the system, although perhaps not for weak localization
phenomena.  Below we consider a specific ring (not an ensemble) and
ask how the maximum amplitude of the persistent current is affected by
zero-point fluctuations.

The effect of a thermal bath on the persistent current has been
discussed by Landauer and B\"uttiker\cite{landauer85} and
B\"uttiker\cite{buettiker:squid,buttiker85} within a Debye relaxation
approach. In this approach, however, the bath affects essentially only
the population of different states, but not the electronic states of
the ring itself. As a consequence, at zero temperature, this model
exhibits no effect due to zero point fluctuations: the magnitude of
the persistent current is independent of the coupling strength to the
bath. Still a different model, introduced by
B\"uttiker,\cite{buettiker:reservoir} considers a ring coupled via a
side branch to a normal electron reservoir. Due to the connection
between reservoir and ring a carrier in the ring eventually escapes to
the reservoir and is replaced with an incident carrier with a phase
which is unrelated to that of the escaping carrier.  This model
predicts even at zero temperature an amplitude of the persistent
current which depends on the coupling
strength.\cite{buettiker:reservoir,ibm88,mello:ringstub,%
moskalets:ringstub,liuchu:ringstub} This effect is not due to
zero-point fluctuations but results from the exchange of carriers
between the reservoir and the ring. If the ring is coupled to a side
branch of finite length, the side branch can nevertheless generate
effects which are similar to that of a reservoir, especially if only
ensemble averaged quantities are considered. This is correct only if
the side branch has a charging energy which is weak compared to the
level spacing.\cite{me:ringstub} If the charging energy is large
compared to the level spacing the side branch has no effect on the
ensemble averaged persistent current.\cite{me:ringstub} It is thus
interesting to ask, whether there exist models which are strictly
canonical (without carrier exchange with a reservoir) and for which
the sample specific persistent current depends nevertheless on the
properties of the bath. The model investigated in this work examines a
ring which is coupled to the bath only via the long range Coulomb
force.

The mesoscopic ring which we consider (see Fig.~\ref{deph:system}) is
divided into two regions by tunneling barriers. It is a ring with an
in-line quantum dot.\cite{buettiker:ringdot:prl,buettiker:ringdot,%
sarma,sand,eckle:kondoring,kang:kondoring,note1}.  This model allows a
simple characterization of the electrostatic potential in terms of
only two variables $U_d$ (for the dot) and $U_a$ (for the arm), and by
the charges $Q_d$ and $Q_a$.  The two regions, the dot and the arm of
the ring, are coupled via capacitors $C_1$ and $C_2$ to the external
circuit.  The external circuit is described by its impedance
$Z_{ext}$.  The potential at the capacitor $C_1$ is denoted by $V_0$,
the charge by $Q_0$.  Likewise, we write the potential and the charge
on the capacitor $C_2$ as $V_\infty$ and $Q_\infty$.

\section{Langevin equation}
\begin{figure}
\centerline{\epsfysize=5cm\epsfbox{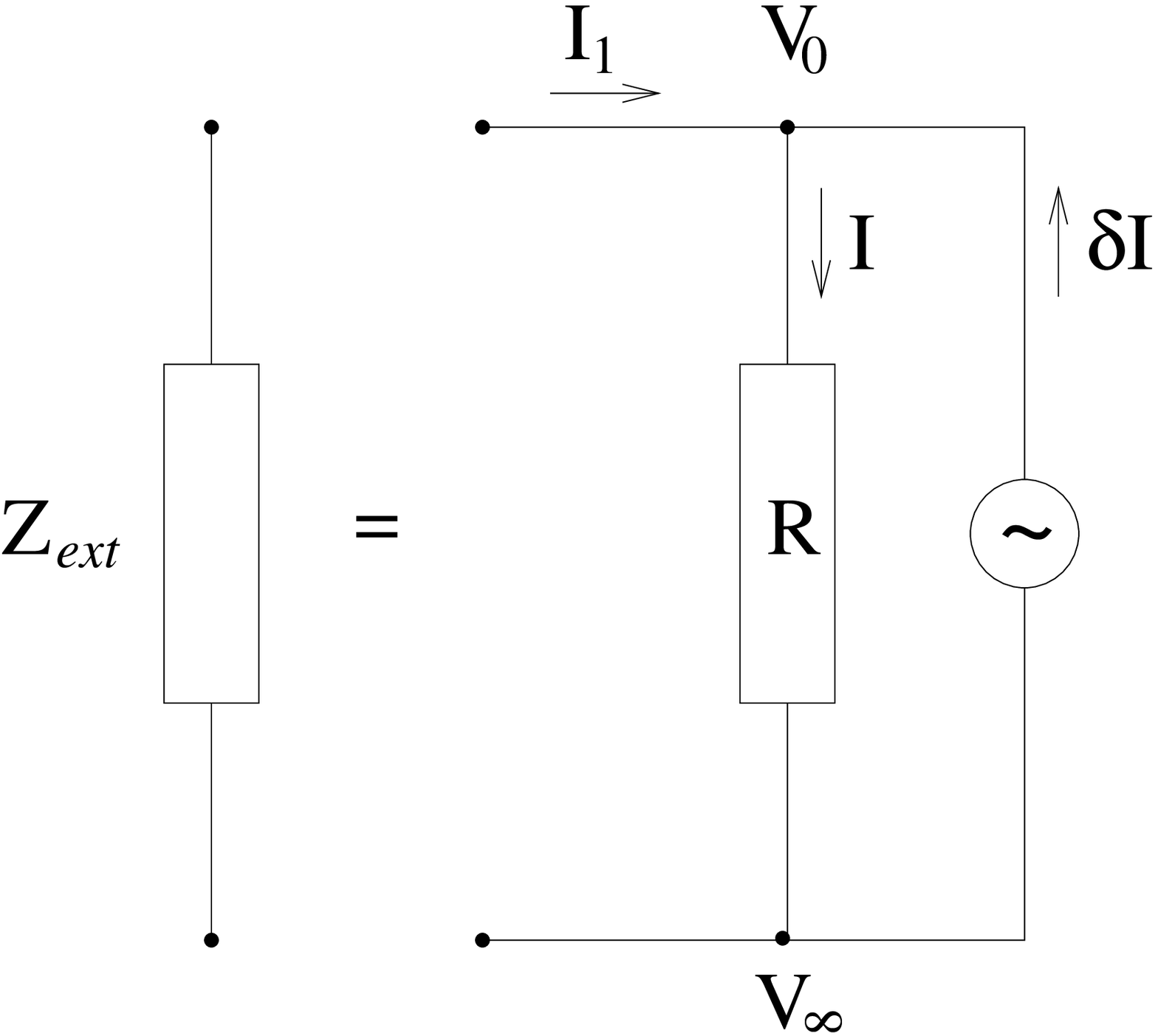}}
\vspace*{0.3cm}
\caption{\label{deph:noise}The external impedance $Z_{ext}(\omega)$
has been replaced by a resistor $R$ with a noise source in parallel in
order to take into account quantum fluctuations.}
\end{figure}

To be specific, we consider a purely resistive external impedance,
$Z_{ext} = R$. The fluctuations generated by this resistor can be
represented as a noise source in parallel with the resistor as shown
in Fig.~\ref{deph:noise}.  The current $\delta V/R$ through the
resistor must be equal to the sum of the current of the noise source
$\delta I(t)$ and of the current through the ring-dot structure. In
terms of the potential difference $\delta V \equiv V_{0} -V_{\infty}$
across the ring-dot structure we find that the coupling between the
ring-dot structure and the external circuit is described by
\begin{equation}
\label{deph:voltagedeq}
{\delta V/R} = - C_0 \delta\dot{V}
+ {C_0 \over C_i} \dot{Q}_d + \delta I(t) .
\end{equation}
Here we introduce the capacitance $C_0^{-1} \equiv C_i^{-1} +
C_e^{-1}$ which is the series capacitance of the {\it internal}
capacitance $C_i \equiv C_L+C_R$ and the {\it external} capacitance
$C_e^{-1} \equiv C_1^{-1} + C_2^{-1}$, and the charge on the dot
$Q_d$. A detailed consideration of the circuit equations leading to
Eq.~(\ref{deph:voltagedeq}) is given in Appendix~\ref{coulomb}.  The
noise source in parallel to the resistor generates a current noise
spectrum, $\langle \delta I(\omega) \delta I (\omega') \rangle \equiv
2\pi \delta(\omega+\omega') S_{II}(\omega)$, given by
\begin{equation}
\label{deph:SII}
S_{II}(\omega) = {\hbar \omega \over R}
\coth \left( {\hbar \omega \over 2kT} \right),
\end{equation}
where $k$ is the Boltzmann constant, and $T$ is the temperature. 
To complete the description of this system, we need to investigate 
the dynamics of the ring in the presence of the fluctuating 
external voltage $\delta V(t)$. 

First, we consider the effect of the external circuit in linear
response.  Since we are only interested in the weak coupling limit, we
can choose the ground state of the ring-dot system in the absence of
an external circuit as a state around which we can expand. What we
need is the linear response relation between $\delta Q_d(t)$ (the
deviation of the charge away from a reference state) as a function of
the external voltage $\delta V(t)$.  Note that both of these variables
are in fact operators.  Here, we proceed as in the random phase
approximation and use a linear response function to describe the
connection between these two quantities. In linear response we can
Fourier transform $\delta Q_d(t)$ and $\delta V(t)$. The linear
response coefficient gives the increment of the charge on the dot in
response to a variation of an external voltage: It is thus a dynamic
capacitance which we denote by $C_d (\omega) = \delta Q_d(\omega) /
\delta V(\omega)$.  Similarly, the relation between the charge $\delta
Q_0$ piled up on the external capacitor $C_1$ and the applied voltage
is determined by the overall capacitance of the ring dot-structure
vis-a-vis the external circuit, and is denoted by $C_{\mu}(\omega) =
\delta Q_0(\omega) / \delta V(\omega)$.  From circuit theory (see
Appendix~\ref{coulomb}) it follows that these two capacitances are
related
\begin{equation}
\label{deph:Cmudef}
C_\mu(\omega) = C_0 - {C_0 \over C_i} C_d(\omega)
\end{equation}
The noise source (see Fig.~\ref{deph:noise}) sees a total impedance
$Z(\omega)$ which consists of this capacitance and the resistance
\begin{equation}
\label{deph:Zt}
{1 \over Z(\omega)}
= \left[ {1 \over R} - i\omega C_\mu(\omega) \right].
\end{equation}
With these response coefficients Eq.~(\ref{deph:voltagedeq}) 
leads to
\begin{equation}
\label{deph:voltagedeqF}
\delta V(\omega) = Z(\omega) \delta I(\omega) = 
\left[1 - i\omega RC_\mu(\omega) \right]^{-1} R \delta I(\omega) 
\end{equation}
and the spectrum of the external voltage is thus 
\begin{equation}
S_{VV}(\omega) = \frac{R {\hbar \omega}
\coth \left( {\hbar \omega \over 2kT} \right)} 
{\left[ 1 + \omega^{2} (RC_\mu(\omega))^{2} \right]}
\label{vspectrum} 
\end{equation}
Our next task is now to find explicit expressions for the 
capacitances $C_{\mu}(\omega)$ and $C_{d}(\omega)$. 

\subsection{The dynamics of the ring}
The dynamics of the charge on the dot, determined by $\hat{Q}_d$, is
in the Heisenberg picture given by
\begin{equation}
{\partial \over \partial t} \hat{Q}_d
= {\hbar \over i} [\hat{H}_{ring}, \hat{Q}_d],
\end{equation}
where $\hat{H}_{ring}$ is the Hamiltonian of the ring in an applied
external potential $\delta V$.

We assume that the tunneling amplitudes are much smaller in magnitude
than the level spacing in the dot and the level spacing in the arm.
Moreover, we assume the electrons to be spinless.  As discussed in
Refs.~\onlinecite{buettiker:ringdot:prl,buettiker:ringdot} a spin
singlet appears in the case of electrons with spin, and the tunneling
amplitude is enhanced by a factor of $\sqrt{2}$.  Our choice of
parameters excludes the Kondo effect,\cite{eckle:kondoring} also if
the electron spin is taken into account.  As a matter of fact, the
Kondo effect appears only at a tunneling amplitude comparable to or
larger than the mean level spacing.\cite{kang:kondoring} We note that
the number of charge carriers in the ring is conserved.  Following
B\"uttiker and Stafford,\cite{buettiker:ringdot:prl,buettiker:ringdot}
we consider hybridization between the topmost occupied electron level
in the arm and the lowest unoccupied electron level in the dot,
$\epsilon_{aM}$ and $\epsilon_{d(N+1)}$ only.  To simplify the
notation, we denote the tunneling amplitudes between the levels
$\epsilon_{aM}$ and $\epsilon_{d(N+1)}$ by $t_L$ for tunneling through
the left junction and by $t_R$ for tunneling through the right one,
and introduce the total tunneling amplitude
\begin{equation}
\label{deph:Delta0}
{\hbar \Delta_0 \over 2}
= \sqrt{t_L^2 + t_R^2 \pm 2 t_L t_R \cos{2\pi \Phi \over \Phi_0}},
\end{equation}
where $\Phi_0 = hc/e$ is the flux quantum, and the sign in front of
the cosine depends on the parity of the number of electrons in the
ring.  The sign is positive, if the number of electrons is odd, and
negative if the number is even.  In the model introduced above, we are
considering only two states, namely the topmost electron being in the
dot, which we represent by the vector $(1,0)$, and the topmost
electron being in the arm of the ring, written as $(0,1)$.  The
dynamics of these two states is described by a time dependent
Hamiltonian
\begin{equation}
\label{deph:Hring}
\hat{H}_{ring}
= {\hbar \varepsilon(t) \over 2} \sigma_z
- {\hbar \Delta_0 \over 2} \sigma_x
+ {\hbar \nu(t) \over 2} {\bf 1},
\end{equation}
where $\sigma_z$ and $\sigma_x$ are Pauli matrices, and $\bf 1$ is the
identity matrix.  The prefactor $\hbar\nu(t)= -C_0 \, \delta V^2(t)$
is a global shift in energy.  We split up the detuning
$\varepsilon(t)$ into a time independent and a time dependent part
$\varepsilon(t) = \varepsilon_0 + \delta\varepsilon(t)$, with
\begin{eqnarray}
\label{deph:eps0}
\hbar \varepsilon_0
&\equiv& \epsilon_{d(N+1)}-\epsilon_{aM}
+ {e^2 (N-N_+ +1/2) \over C} - {C_e \over C}V_0
\nonumber\\
&\equiv& {e \over C} \left( Q_{d*} - Q_{d0} \right),\\
\label{deph:Deps}
\delta\varepsilon(t)
&\equiv& {e \over \hbar} {C_0 \over C_i} \delta V(t). 
\end{eqnarray}
The effective background charge on the dot is $Q_{d0} =
eN_{+}+{C_e}V_0$, where the first term is a built in background charge
and the second term can be externally controlled by applying a static
voltage across the ring-dot structure.  If coherent tunneling is
neglected the ground states of the ring with $N$ and $N+1$ carriers
are degenerate if the polarization charge is equal to
\begin{equation}
Q_{d*} \equiv e \left( N + {1 \over 2} \right)
+ {C \over e} \left( \epsilon_{d(N+1)} - \epsilon_{aM} \right) . 
\end{equation}
Thus $Q_{d0} = Q_{d*}$ is the (classical) condition at which the
Coulomb blockade is lifted. Quantum mechanically the state in the ring
and the state in the arm of the ring are in resonance when $Q_{d0} =
Q_{d*}$ and the persistent current exhibits a peak.  Our simple
two-level picture is applicable when $\varepsilon_0$ is small, that
is, in the vicinity of a resonance.

We want to find the time evolution of a state $\psi(t)$,
which we write as
\begin{equation}
\label{deph:state}
\psi = e^{i\chi/2}
{\cos{\theta \over 2} \, e^{i\varphi/2} \hfill
\choose
\sin{\theta \over 2} \, e^{-i\varphi/2}},
\end{equation}
with $\theta$, $\varphi$ and $\chi$ real.  This is the most general
form of a normalized complex vector in two dimensions.  In terms of
$\theta$, $\varphi$ and the global phase $\chi$, the time dependent
Schr\"odinger equation reads
\begin{eqnarray}
\label{deph:central1}
\dot{\varphi} &=& -\varepsilon_{0} - \delta \varepsilon (t) 
- \Delta_0 \cot\theta \cos\varphi, \\
\label{deph:central2}
\dot{\theta} &=& -\Delta_0 \sin\varphi, \\
\label{deph:chideq}
\dot{\chi} &=& -\nu + \Delta_0 {\cos\varphi \over \sin\theta}.
\end{eqnarray}
Note that Eq.~(\ref{deph:chideq}) describing the dynamics of the
global phase $\chi$ is special in the sense that it is driven by
$\phi$ and $\theta$ but it has no
back-effect onto them.  Moreover, $\chi$ is irrelevant for expectation
values, like the persistent current or the charge on the dot.  It
plays an important role, however, in the discussion of phase diffusion times, 
see
Sec.~\ref{deph:dephasing}.  The system of the two equations for
$\varphi$ and $\theta$,
Eqs.~(\ref{deph:central1},~\ref{deph:central2}), is closed by
Eq.~(\ref{deph:voltagedeq}),
\begin{equation}
\dot{Q}_d = {d \over dt} \langle \psi(t) | \hat{Q}_d
| \psi(t) \rangle,
\end{equation} 
with $\psi(t)$ defined above, and
\begin{equation}
\hat{Q}_d = e \left( \matrix{ 1 & 0 \cr 0 & 0 } \right)
+ e N_+,
\end{equation}
We find 
\begin{equation}
\label{deph:dQd}
\dot{Q}_d = e {d \over dt}
\langle \psi(t) |
\left( \matrix{ 1 & 0 \cr 0 & 0 } \right)
| \psi(t) \rangle
= -{e \over 2} \sin\theta \, \dot{\theta}.
\end{equation}
Using this result and Eq.~(\ref{deph:voltagedeq}),
gives 
\begin{equation}
\label{deph:central3}
\delta V = - C_0 R \delta\dot{V} -
{C_0 \over C_i} {e \over 2} \sin\theta \, \dot{\theta}
+ R \delta I(t) .
\end{equation}
Eqs.~(\ref{deph:central1},~\ref{deph:central2}) and
(\ref{deph:central3}) form a closed system of equations in which the
external circuit is incorporated in terms of a fluctuating current
$\delta I (t)$ and of an ohmic resistor $R$. In the next section, we
investigate Eqs.~(\ref{deph:central1},~\ref{deph:central2}) and
(\ref{deph:central3}) to find the effect of zero-point fluctuations on
the persistent current of the ring.

\subsection{Expansion around a stationary state}
\label{deph:expansion}
First, let us discuss the stationary states of the system of
differential equations, Eqs.~(\ref{deph:central1},
\ref{deph:central2}) and (\ref{deph:central3}) in the absence of the
noise term $\delta I(t)$.  We take $0 \le \varphi < 2\pi$ and $0 \le
\theta < \pi$.  This gives immediately $\sin\varphi = 0$ and
consequently a stationary state has $\varphi \equiv \varphi_0$, with
$\varphi_0 = 0$ or $\varphi_0 = \pi$. With this it is easy to show
that in the stationary state we must have $\theta \equiv \theta_0$,
with
\begin{equation}
\cot\theta_0 = \pm {\varepsilon_0 \over \Delta_0}. 
\end{equation}
The lower sign applies for $\varphi_0 = 0$.  This is the ground state
for the ring-dot system at fixed $\varepsilon(t) \equiv
\varepsilon_0$, and the upper sign holds for $\varphi_0 = \pi$.
The energy of the ground state is $-\hbar\Omega_0/2$, thus the global
phase is $\chi_0(t) = \Omega_0 t$. 
Here 
\begin{equation}
\Omega_0^2 \equiv \varepsilon_0^2 + \Delta_0^2 
\end{equation}
is the resonance frequency of the (decoupled) two-level system. 
We also introduce 
the ``classical'' relaxation time $\tau_{RC} \equiv RC_0$, and a
relaxation rate
\begin{equation}
\label{deph:Gamma}
\Gamma \equiv \pi \alpha {\Delta_0^2 \over \Omega_0}
\end{equation}
which as 
we shall see in Sec.~\ref{deph:weakcoupling} 
is a relaxation rate due to the coupling of the ring-dot system 
to the external circuit. Here $\alpha$ is a dimensionless 
coupling constant 
\begin{equation}
\label{deph:alphasimple}
\alpha \equiv {R \over R_K} \left( {C_0 \over C_i} \right)^2
\end{equation}
with $R_K \equiv h/e^2$ the quantum of resistance.  We will
immediately see the usefulness of these definitions.

Now, we switch on the noise $\delta I(t)$.  We seek $\varphi(t) ,
\theta(t)$, $\chi(t)$ and $\delta V(t)$ in linear order in the noise
current $\delta I(t)$.  We expand $\varphi(t)$ and $\theta(t)$ to
first order around the ground state, $\varphi = 0$ and $\theta =
\theta_0$. For $ \delta\varphi(t) = \varphi(t) - \varphi_0$,
$\delta\theta(t) = \theta (t) - \theta_0$, we find in Fourier space,
\begin{eqnarray}
\label{deph:central1linF}
-i \omega \delta\varphi &=& -\delta\varepsilon
+ {\Omega_0^2 \over \Delta_0} \delta\theta, \\
\label{deph:central2linF}
-i \omega \delta\theta &=& -\Delta_0 \delta\varphi, \\
\label{deph:central3linF}
-i \omega \delta \varepsilon
&=& {1 \over \tau_{RC}} \left[
-\delta\varepsilon - \Gamma \delta\varphi
+ {e \over \hbar} R {C_0 \over C_i} \delta I
\right]. 
\end{eqnarray}
We also expand the global phase $\chi(t)$ around its
evolution in the ground state $\chi_0(t) = \Omega_0 t$, and define
$\delta\chi(t) = \chi(t) - \chi_0(t)$.  In Fourier space,
Eq.~(\ref{deph:chideq}) becomes
\begin{equation}
\label{deph:chideqF}
-i\omega \delta\chi = \Omega_0 {\varepsilon_0 \over \Delta_0}
\delta\theta.
\end{equation}
We note that there is no effect of the global shift in energy,
$\hbar\nu(t)$, as it is quadratic in the voltage $\delta V$, and we
are only interested in effects up to linear order in $\delta V$.

In the following section, we evaluate the linear response of the ring,
described by Eqs.~(\ref{deph:central1linF},~\ref{deph:central2linF})
to an applied external potential $\delta V(\omega)$, giving the
frequency dependent capacitance $C_\mu(\omega)$.

\subsection{Capacitance of the ring and impedance}
We evaluate $\delta Q_d(\omega)$ to first order in $\delta V(\omega)$,
using Eqs.~(\ref{deph:central1linF},~\ref{deph:central2linF}), which
gives the dynamic capacitance $C_d(\omega) = \delta Q_d(\omega)/\delta
V(\omega)$,
\begin{equation}
\label{deph:Cd}
C_d(\omega) = C_0 {e^2/(2C_i) \over \hbar\Omega_0}
{\Delta_0^2 \over \omega^2 - \Omega_0^2}.
\end{equation}
The frequency dependent capacitance as seen from the external circuit
reads, see Eq.~(\ref{deph:Cmudef}),
\begin{equation}
\label{deph:Cmu}
C_\mu(\omega) = C_0 \left(
1 - {e^2/(2C_i) \over \hbar\Omega_0}
{\Delta_0^2 \over \omega^2 - \Omega_0^2}
\right).
\end{equation}
The frequency dependent capacitance $C_{\mu}$ contains in addition to
the geometrical capacitance $C_0$ a term which arises from the dynamic
polarizability of the ring.  In the low frequency regime the
polarizability enhances the capacitance,
\end{multicols}
\widetext
\vspace*{-0.2truein} \noindent \hrulefill \hspace*{3.6truein}
\begin{equation}
\label{deph:Cmuzero}
C_\mu(0) = C_0 \left(
1 + {e^2/(2C_i) \over \hbar\Omega_0}
{\Delta_0^2 \over \Omega_0^2}
\right)
= C_0 + \left( {C_0 \over C_i} \right)^2
{2e^2 |t_\pm|^2
\over
\left\{
\left[ e(Q_{d0} - Q_{d*})/C \right]^2 + 4|t_\pm|^2
\right\}^{3/2}}.
\end{equation}
\hspace*{3.6truein} \noindent \hrulefill
\begin{multicols}{2}
\narrowtext
\noindent
The zero frequency capacitance has already been obtained by Stafford
and one of the authors.\cite{buettiker:ringdot:prl,buettiker:ringdot}
The zero-frequency (electro-chemical) capacitance
Eq.~(\ref{deph:Cmuzero}) can be obtained as a second order derivative
of the grand canonical potential\cite{buettiker:ringdot} with respect
to the voltage applied across the ring dot-system or as a dynamic
response to a slowly varying voltage across the ring-dot system.
Since the tunneling amplitude $\Delta_0$ depends periodically on the
magnetic flux $\Phi$, see Eq.~(\ref{deph:Delta0}), $C_\mu(\omega)$ can
be modulated by varying $\Phi$.  The electro-chemical capacitance
exhibits a peak as the bias $V_0$ is varied.  The peak shows up at
resonance, $Q_{d0} = Q_{d*}$, corresponding to $\varepsilon_0 = 0$.
At the same point, there is also a peak in the persistent current. A
recent experiment by Deblock {\em et al.}\cite{deblock:screening}\ 
demonstrated indeed a flux dependent polarizability of mesoscopic
rings.

At high frequencies the polarization cannot follow the external
voltage and the capacitance Eq.~(\ref{deph:Cmu}) is smaller than the
geometrical capacitance. At $\omega = \Omega_{0}$ the capacitance
exhibits a resonance characteristic for dielectric functions.  The
ring by itself has no damping mechanism, therefore the resonance shows
up as a pole.  The damping is provided by the dissipation in the
external circuit.

From Eqs.~(\ref{deph:Zt}) and (\ref{deph:Cmu}), we find
\begin{equation}
{1 \over Z(\omega)} = {1 \over R}
{\omega^2 - \Omega_0^2
- i\omega \tau_{RC} \left( \omega^2 - \Omega_0^2
- \Gamma/\tau_{RC} \right)
\over \omega^2 - \Omega_0^2}.
\end{equation}
This expression for the impedance 
seen by the noise source contains all the information needed
to calculate the various spectral densities.

\subsection{Spectral densities}
\begin{figure}[ht]
\centerline{\epsfysize=6.5cm\epsfbox{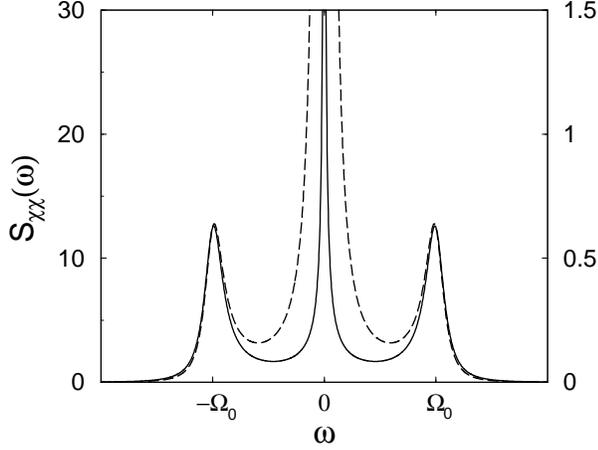}}
\caption{\label{deph:fig:Scc}The spectral function
$S_{\chi\chi}(\omega)$ at finite temperature $kT=10\hbar\Omega_0$
(solid line) and at zero temperature (dashed line).  The left $y$-axis
carries the scale for the finite temperature curve, the right one
carries the scale for the zero temperature curve.  The parameters are
the same as in Fig.~\ref{deph:fig:Spp}, in addition $\varepsilon_0 =
\Omega_0/\sqrt{2}$.}
\end{figure}

We solve Eqs.~(\ref{deph:central1linF})--(\ref{deph:central3linF}) for
$\delta\varphi$ by eliminating $\delta\theta$ and $\delta\varepsilon$,
and obtain after some algebra
\begin{equation}
\delta \varphi = {e \over \hbar} {C_0 \over C_i}
{(-i\omega) Z(\omega) \over \omega^2 - \Omega_0^2} \delta I.
\end{equation}
It follows immediately that $\langle \delta\varphi(\omega) \rangle =
\langle \delta\theta(\omega) \rangle = 0$ and that $\langle
\delta\varepsilon(\omega) \rangle = 0$.  The spectral densities of
$\delta\varphi(\omega)$, $\delta\theta(\omega)$ and their
cross-correlations can all be expressed in terms of the current noise
spectral density.  We have $S_{\theta\theta}(\omega) = (\Delta_0^2 /
\omega^2) S_{\varphi\varphi}(\omega)$, $S_{\varphi\theta}(\omega) = -i
(\Delta_0 / \omega) S_{\varphi\varphi}(\omega)$, and
\begin{eqnarray}
\label{deph:Sphiphi}
S_{\varphi\varphi}(\omega)
= {e^2 \over \hbar^2} \left( {C_0 \over C_i} \right)^2
{\omega^2 |Z(\omega)|^2
\over \left( \omega^2 - \Omega_0^2 \right)^2} S_{II}(\omega).
\end{eqnarray}
We will relate the reduction of the persistent current to
$S_{\varphi\varphi}(\omega)$ in Sec.~\ref{deph:pc}.  While the
spectral densities for $\delta\varphi$ and $\delta\theta$ are finite
at zero frequency, the spectral density of the global phase
$\delta\chi$ is not.  We have $S_{\chi\chi}(\omega) = (\Omega_0^2
\varepsilon_0^2 / \omega^4) S_{\varphi\varphi}(\omega)$, see
Eq.~(\ref{deph:chideqF}), therefore it exhibits a $\omega^{-2}$-pole
in the vicinity of $\omega=0$ for finite temperatures
\begin{equation}
\label{deph:Schichif}
S_{\chi\chi}(\omega)
\sim 2\pi\alpha {2 kT \over \hbar} 
{\varepsilon_0^2 \over \Omega_0^2} {1 \over \omega^2}, \quad
(\omega \to 0),
\end{equation}
whereas in the quantum limit $T=0$ the order of the pole is reduced by
one,
\begin{equation}
\label{deph:Schichiq}
S_{\chi\chi}(\omega)
\sim 2\pi\alpha
{\varepsilon_0^2 \over \Omega_0^2} {1 \over |\omega|}, \quad
(\omega \to 0),
\end{equation}
see Fig.~\ref{deph:fig:Scc}.  These two last results do not affect the
persistent current, but they are of great importance in the discussion
of the rates of phase diffusion in Sec.~\ref{deph:dephasing}.

\subsection{Weak coupling}
\label{deph:weakcoupling}
Within the approach out-lined above, we can expect to reproduce the
exact result for the reduction of the persistent current only in the
limit of weak coupling between the ring-dot system and the external
circuit.  This means that the {\em potential} fluctuations of the
system and the external circuit are only weakly coupled. Thus weak
coupling between the ring and the external circuit is achieved by
letting the capacitances coupling the two systems become very
small,\cite{note2} $C_1,\, C_2 \to 0$, for a fixed resistance $R$.
This entails $C_0 \to 0$, thus the coupling strength $\alpha$ between
the ring and the external circuit, Eq.~(\ref{deph:alphasimple}), is
small against 1.  Note that in this approximation the RC-time
$\tau_{RC}$ becomes very small as well.

\begin{figure}[ht]
\centerline{\epsfysize=6.5cm\epsfbox{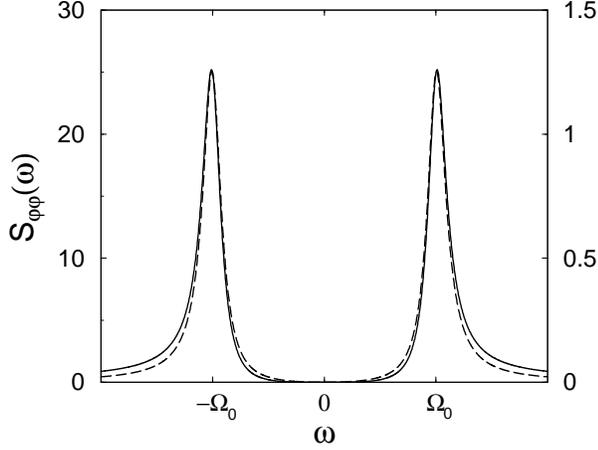}}
\caption{\label{deph:fig:Spp}The spectral function
$S_{\varphi\varphi}(\omega)$ at finite temperature
$kT=10\hbar\Omega_0$ (solid line) and at zero temperature (dashed
line).  The left $y$-axis carries the scale for the finite temperature
curve, the right one carries the scale for the zero temperature curve.
The parameters are $R=5R_K$, $C_0=0.1C_i$, $\Delta_0=0.5\Omega_0$ and
$\Gamma=0.2\Omega_0$.}
\end{figure}

In the following paragraph, we calculate the poles of $Z(\omega)$, in
the approximation $C_0 \to 0$. The small parameters in
this case are $\tau_{RC} \propto C_0$ and $\Gamma \propto C_0^2$.  The
equation $[Z(\omega)]^{-1}=0$ has one solution behaving as $C_0^{-1}$.
To leading order in $C_0$, it reads
\begin{equation}
\omega_3 = -{i \over \tau_{RC}}.
\end{equation}
Thus in the weak coupling limit charge relaxation across the mesoscopic 
system becomes instantaneous. 
The other two solutions, $\omega_\pm$, are of zeroth order in $C_0$.
Up to and including corrections of order $C_0^2$, they are 
\begin{equation}
\omega_\pm = \pm \Omega_0 - i{\Gamma \over 2},
\end{equation}
with a relaxation rate $\Gamma$ introduced in Eq.~(\ref{deph:Gamma}).
The rate $\Gamma$ describes the relaxation of perturbations with a
frequency close to the eigenfrequency of the isolated ring-dot system.
In Sec.~\ref{deph:dephasing} it is shown that $\Gamma$ is also
related to a phase diffusion time.  The denominator in
Eq.~(\ref{deph:Sphiphi}) can now be written as
\end{multicols}
\widetext
\vspace*{-0.2truein} \noindent \hrulefill \hspace*{3.6truein}
\begin{equation}
{\left( \omega^2 - \Omega_0^2 \right)^2
\over |Z(\omega)|^2}
= {\tau_{RC}^2 \over R^2}
\left( \omega^2 + {1 \over \tau_{RC}^2} \right)
\left[
\left( \omega - \Omega_0 \right)^2
+ \left( \Gamma / 2 \right)^2
\right]
\left[
\left( \omega + \Omega_0 \right)^2
+ \left( \Gamma / 2 \right)^2
\right].
\end{equation}
For frequencies, $\omega \tau_{RC} \ll 1$, the denominator takes
the form
\begin{equation}
\label{deph:essential} 
{\left( \omega^2 - \Omega_0^2 \right)^2
\over |Z(\omega)|^2}
= {1 \over R^2} \left[
\left( \omega - \Omega_0 \right)^2
+ \left( \Gamma / 2 \right)^2
\right]
\left[
\left( \omega + \Omega_0 \right)^2
+ \left( \Gamma / 2 \right)^2
\right] . 
\end{equation}
Since in the quantum limit it is always necessary to introduce a
cut-off, Eq.~(\ref{deph:essential}) is valid over the entire range of
frequencies of interest, if only $1/\tau_{RC}$ is larger than the
cut-off frequency.  The spectral density $S_{\varphi\varphi}(\omega)$,
Eq.~(\ref{deph:Sphiphi}), reads therefore
\begin{equation}
S_{\varphi\varphi} (\omega)
= {2\pi\alpha\omega^3 \coth {\hbar\omega \over 2kT}
\over
\left[
\left( \omega - \Omega_0 \right)^2
+ \left( \Gamma / 2 \right)^2
\right]
\left[
\left( \omega + \Omega_0 \right)^2
+ \left( \Gamma / 2 \right)^2
\right]},
\end{equation}
\hspace*{3.6truein} \noindent \hrulefill
\begin{multicols}{2}
\narrowtext
\noindent
where the coupling strength $\alpha$ between the ring and the external
circuit is defined in Eq.~(\ref{deph:alphasimple}).  Note that
$S_{\varphi\varphi}(\omega)$ goes to zero at the origin $\omega=0$
like $|\omega|^3$ in the extreme quantum case ($T=0$), and like
$\omega^2$ for finite temperatures, see Fig.~\ref{deph:fig:Spp}.  For
$|\omega| \gg \Omega_0$, it behaves like $|\omega|^{-1}$ for small
temperatures and like $\omega^{-2}$ for large temperatures.  Analogous
formul\ae\ hold for $S_{\theta\theta}(\omega)$ and for
$S_{\varphi\theta}(\omega)$.  We mention that the spectral density
$S_{\theta\theta}(\omega)$ behaves for small $\omega$ like $|\omega|$
at $T=0$ and goes to a constant for finite $T$.  All the three
spectral densities exhibit peaks of width $\Gamma$ at $\pm \Omega_0$.

The spectral density of the global phase $S_{\chi\chi}(\omega)$,
Fig.~\ref{deph:fig:Scc}, exhibits peaks at $\omega = \pm \Omega_0$ as
well.  More important, however, is the fact that it has a pole at
$\omega=0$, as already pointed out in the previous section.

\section{Persistent current}
\label{deph:pc}
In the following, we calculate the persistent current as a quantum 
{\it and} statistical average of the operator of
the circulating current
\begin{equation}
\hat{I}_c = \left(
\matrix{ 0 & {\cal J} \cr {\cal J}^* & 0 }
\right),
\label{icop}
\end{equation}
where $\cal J$ is given by
\end{multicols}
\widetext
\vspace*{-0.2truein} \noindent \hrulefill \hspace*{3.6truein}
\begin{equation}
\label{deph:pcurop}
{\cal J} = -{e \over \hbar}
\left( t_L^2 + t_R^2
\pm 2t_L t_R \cos {2\pi\Phi \over \Phi_0} \right)^{-1/2}
\left\{ \pm t_L t_R \sin {2\pi\Phi \over \Phi_0}
-i\left[
{C_R \over C_i} t_L^2 - {C_L \over C_i} t_R^2
\pm {C_R - C_L \over C_i} t_L t_R \cos {2\pi\Phi \over \Phi_0}
\right] \right\}.
\end{equation}
\hspace*{3.6truein} \noindent \hrulefill
\begin{multicols}{2}
\narrowtext
\noindent
Here the first term is a pure particle current contribution, whereas
the second term is a consequence of interactions.  For the average
persistent current of interest here, it is only the first term which
contributes. A derivation of Eq.~(\ref{deph:pcurop}) and a discussion
of the relationship between $I_c$ and the magnetization is given in
Appendix~\ref{circ}.

The expectation value of the persistent current for the state given in
Eq.~(\ref{deph:state}) reads
\begin{equation}
I(t) \equiv \langle \psi(t) | \hat{I}_c | \psi(t) \rangle
= {1 \over 2} {\rm Re}\, {\cal J} \sin\theta \, e^{-i\varphi}.
\end{equation}
We are, however, interested in the {\em statistically averaged\/} 
of the persistent
current $\langle I(t) \rangle$, given by
\begin{equation}
\langle \langle \psi(t) | \hat{I}_c | \psi(t) \rangle \rangle
= {1 \over 2} {\rm Re}\, \left( {\cal J}
\langle \sin\theta \, e^{-i\varphi} \rangle \right), 
\end{equation}
where the double bracket indicates a quantum and statistical average. 
Therefore, we have to calculate the correlator $\langle \sin\theta \,
e^{-i\varphi} \rangle$.  First, we observe that there are no
correlations between $\delta\varphi$ and $\delta\theta$, 
\begin{equation}
\langle \delta\varphi(t) \delta\theta(t) \rangle = 0,
\end{equation}
as the spectral density $S_{\varphi\theta}(\omega)$ is an odd function
of $\omega$.  Thus we have
\begin{equation}
\langle \sin\theta e^{-i\varphi} \rangle
= \langle \sin\theta \rangle
\langle e^{-i\varphi} \rangle,
\end{equation}
to second order in $\delta I$, that is, the probability distributions
for $\varphi$ and $\theta$ are decoupled up to second order in $\delta
I$.  Second, it can be shown that the second order corrections to
$\varphi$ and $\theta$ vanish on the average.  Finally, we assume that
the correlations in $\delta\varphi$ are Gaussian, which allows us to
write
\begin{equation}
\langle e^{-i\varphi(t)} \rangle
= e^{-i\varphi_0} \langle e^{-i\delta\varphi(t)} \rangle
= \left\langle
\exp\left( -{\delta\varphi^2(t) \over 2} \right)
\right\rangle,
\end{equation}
where we have used that $\varphi_0 = 0$.  In the weak coupling limit,
see Sec.~\ref{deph:weakcoupling}, and in the extreme quantum limit,
$T=0$, we find for the time averaged mean square fluctuations
\begin{equation}
\label{deph:msdphi2}
\langle \delta\varphi^2(t) \rangle
= \int_0^{\omega_c} {d\omega \over \pi}
S_{\varphi\varphi}(\omega)
\approx 2\alpha
\ln {\omega_c \over \Omega_0},
\end{equation}
where the cut-off frequency $\omega_c$ is taken to be larger than all
frequency scales (except $\tau_{RC}^{-1}$), and $\alpha$ is the
coupling strength between ring and external circuit, see
Eq.~(\ref{deph:alphasimple}). In the limit $\omega_c \gg \Omega_0$, we
can neglect $\langle \delta\theta^2(t) \rangle = \int_0^{\omega_c}
(d\omega/\pi) S_{\theta\theta}(\omega)$ against $\langle
\delta\varphi^2(t) \rangle$.  We insert $\langle \delta\varphi^2(t)
\rangle$ and $\sin\theta_0 = \Delta_0/\Omega_0$ into $\langle
\sin\theta \, e^{-i\varphi} \rangle$, and observe that
\begin{equation}
{1 \over 2} {\rm Re}\, {\cal J}
= {\hbar c \over 2} {\partial \Delta_0 \over \partial \Phi}.
\end{equation}
Put together, we obtain the noise averaged current in the ring
\begin{equation}
\label{deph:pcsimple}
\langle I(t) \rangle
= -{\hbar c \over 2} {\partial \Delta_0 \over \partial \Phi}
{\Delta_0 \over \Omega_0}
\left( {\Omega_0 \over \omega_c} \right)^\alpha.
\end{equation}
Eq.~(\ref{deph:pcsimple}) is a key result of this work.  For $\alpha
\ll 1$, corresponding to weak coupling between the ring and the
external circuit, the power law for the persistent current obtained in
Eq.~(\ref{deph:pcsimple}), as well as the exponent $\alpha$,
Eq.~(\ref{deph:alphasimple}), are the same as the one obtained when
the external circuit is treated quantum mechanically as well.  In a
recent work,\cite{cedraschi:prl} the authors in collaboration with
Ponomarenko have shown that if the external circuit is represented by
a transmission line, the persistent current at resonance, $\varepsilon
= 0$, has for $\alpha < 1$ the power law behavior
\begin{equation}
\label{deph:pcvpmb}
I(\varepsilon=0) \propto
\left(
{\Delta_0 \over \omega_c}
\right)^{\alpha \over 1-\alpha}.
\end{equation}
For a very small coupling parameter, $\alpha \ll 1$, the Bethe ansatz
result Eq.~(\ref{deph:pcvpmb}), goes over to the power law of
Eq.~(\ref{deph:pcsimple}).  Thus the simplified discussion presented
here leads at least in the weak coupling limit to the same result as
the one obtained in Ref.~\onlinecite{cedraschi:prl}.

We emphasize that the persistent current is a property of the ground
state of a system. In our case, the persistent current is, however,
carried by only a part of the system.  Due to the coupling to the
external circuit this subsystem is subject to fluctuations which even
at zero temperatures suppress the persistent current.  If we keep the
capacitances fixed, Eqs.~(\ref{deph:pcsimple}) and (\ref{deph:pcvpmb})
give a persistent current which decreases with increasing external
resistance $R$.  The spectral density of the voltage fluctuations
across the ring dot system increases with $R$ (see
Eq.~(\ref{vspectrum})).  We next characterize the fluctuations of the
ring-dot subsystem in more detail.

\section{Phase Diffusion Times}
\label{deph:dephasing}
In this work we are concerned with the effect of an external circuit
on the ground state properties of the system and in particular on the
persistent current of the ring-dot subsystem. In transport
experiments which reveal phase-coherent contributions or in
experiments which investigate the evolution of an initial state, we
can describe the effect of an environment with the help of dephasing
rates. On the other hand, it is not a priori clear that the coherence
properties of the ground state of a system, can be equally expressed
in terms of dephasing rates. After all the persistent current
investigated here exists quite independently on how long we let the
system evolve. Here we consider the state in the ring and investigate
its evolution away from an initial state.  Below we show that this
evolution is diffusive at least for times short enough such that we
remain within the limit of validity of the calculation.

We have seen in Sec.~\ref{deph:weakcoupling} that the dynamics of the
ring strongly affects the spectral densities
$S_{\varphi\varphi}(\omega)$ and $S_{\theta\theta}(\omega)$ in the
vicinity of the characteristic frequency $\Omega_0$.  The
characteristic frequency describes the free dynamics of a ring which
is disconnected from the external circuit, described by the
Hamiltonian $\hat{H}_{ring}$, Eq.~(\ref{deph:Hring}), without the time
dependent terms $\delta\varepsilon(t)$ and $\nu(t)$,
\begin{equation}
\hat{H}_0 = {\hbar\varepsilon_0 \over 2} \sigma_z
- {\hbar\Delta_0 \over 2} \sigma_x.
\end{equation}
This Hamiltonian can be viewed as describing a spin in a magnetic
field of strength proportional to $(\varepsilon_0^2 +
\Delta_0^2)^{1/2} = \Omega_0$, forming an angle $\theta_0$ with the
$z$-axis.  The angle $\theta_0$ has been defined in
Sec.~\ref{deph:expansion} and is related to $\varepsilon_0$ and
$\Delta_0$ by $\cot\theta_0 = -\varepsilon_0 / \Delta_0$.  The
eigenstates of the time independent Hamiltonian $\hat{H}_0$ are the
{\em ground state\/} $\psi_- = (\cos\theta_0/2, \sin\theta_0/2)$, with
eigenvalue $-\hbar\Omega_0/2$, and the {\em excited state\/} $\psi_+ =
(-\sin\theta_0/2, \cos\theta_0/2)$ with eigenvalue $\hbar\Omega_0/2$.
The evolution due to $\hat{H}_0$ of a spin prepared in a state which
is not an eigenstate corresponds to a rotation of the expectation
value of the spin about the direction of the magnetic field with a
frequency $\Omega_0$. Let us now return to the full problem in which
polarization fluctuations modify the free evolution of the decoupled
system. In the weak coupling limit, we are interested in the time
evolution which is long compared to $\Omega^{-1}_0$.  Therefore, we
switch to the ``rotating frame'' picture, where the wave function
$\psi(t)$ of the ring, Eq.~(\ref{deph:state}), reads $\psi_R(t) \equiv
\exp(i\hat{H}_0 t/\hbar) \psi(t)$.  We consider the projection
$c_\pm(t) \equiv \langle \psi_\pm | \psi_R(t) \rangle$ of the wave
function $\psi_R(t)$ onto the states $\psi_\pm$.  If the state
$\psi(t)$ evolves under the influence of the time independent
Hamiltonian $\hat{H}_0$, the projections $c_\pm(t)$ are constant in
time.  The moduli of the projections, averaged over the noise,
$\langle |c_\pm(t)|^2 \rangle$ are also independent of time if the
evolution is determined by $\hat{H}_{ring}$ which includes the
fluctuating potential.  The phase of the projections $c_\pm(t)$,
however, shows diffusive behavior.  On the average, we have $\langle
|c_\pm(t)-c_\pm(0)|^2 \rangle \sim t/\tau_\pm$ for sufficiently long
times $t$. We shall see below that $\tau_-$ is related to the phase
diffusion of the {\em global\/} phase $\chi$, whereas $\tau_+$ is
related to the diffusion of the {\em internal\/} phases $\varphi$ and
$\theta$, see Eq.~(\ref{deph:state}).  It is a known feature of
two-level systems that they exhibit two distinct dephasing
times.\cite{tian:qubit} The difference between $\tau_-$ and $\tau_+$
is particularly pronounced in the low temperature limit.  The phase
breaking time $\tau_-$ diverges at zero temperature, whereas the time
$\tau_+$ saturates to a finite value for temperatures $kT$ below
$\hbar\Omega_0$.

We expand the wave function $\psi_R$ to first order in
$\delta\varepsilon$, that is, to first order in $\delta\varphi$,
$\delta\theta$ and $\delta\chi$,
\end{multicols}
\widetext
\vspace*{-0.2truein} \noindent \hrulefill \hspace*{3.6truein}
\begin{equation}
\psi_R(t) = \left[
1 + i\left(
{\delta\chi(t) \over 2}
- {\varepsilon_0 \over \Omega_0} {\delta\varphi(t) \over 2}
\right)
\right] \psi_-
+ \left[
{\delta\theta(t) \over 2}
- i{\Delta_0 \over \Omega_0} {\delta\varphi(t) \over 2}
\right] e^{i\Omega_0 t} \psi_+ .
\end{equation}
From this expression, we immediately obtain the projections $c_-(t) =
1 + i[\delta\chi(t)/2 - (\varepsilon_0/\Omega_0) \delta\varphi(t)/2]$
and $c_+(t) = 1/2 (\delta\theta(t) - i(\Delta_0/\Omega_0)
\delta\varphi(t) )$ $\exp(i\Omega_0 t)$.  As the averages $\langle
\delta\varphi^2(t) \rangle$, $\langle \delta\theta^2(t) \rangle$ and
$\langle \delta\chi^2(t) \rangle$ as well as all the ``crossed''
averages of the type $\langle \delta\varphi(t) \delta\theta(t)
\rangle$ are constant in $t$, the moduli $\langle |c_\pm(t)|^2
\rangle$ are indeed independent of time.  The mean squared
displacement of the projections $c_\pm(t)$ read respectively
\begin{eqnarray}
\label{deph:cminus}
\left\langle \left| c_-(t) - c_-(0) \right|^2 \right\rangle
&=& \int {d\omega \over 2\pi} \sin^2 {\omega t\over 2}
\left[
S_{\chi\chi}(\omega)
+ {\varepsilon_0^2 \over \Omega_0^2} S_{\varphi\varphi}(\omega)
\right], \\
\label{deph:cplus}
\left\langle \left| c_+(t) - c_+(0) \right|^2 \right\rangle
&=& \int {d\omega \over 2\pi} \sin^2 {\omega t \over 2}
\left[
S_{\theta\theta}(\omega+\Omega_0)
+ {\Delta_0^2 \over \Omega_0^2} S_{\varphi\varphi}(\omega+\Omega_0)
\right].
\end{eqnarray}
\hspace*{3.6truein} \noindent \hrulefill
\begin{multicols}{2}
\narrowtext
\noindent
The long time behavior of Eq.~(\ref{deph:cminus}) is dominated by the
frequencies near $\omega=0$.  The spectral density
$S_{\varphi\varphi}(\omega)$ vanishes like $\omega^2$ for finite
temperatures or even like $|\omega|^3$ in the zero temperature limit.
The spectral density $S_{\chi\chi}(\omega)$, however has a pole at
$\omega=0$.  For finite temperatures the pole is of order
$\omega^{-2}$, see Eq.~(\ref{deph:Schichif}), entailing a long time
behavior of the type $\langle |c_-(t) - c_-(0)|^2 \rangle \sim
t/\tau_-$, with a dephasing time
\begin{equation}
\label{deph:tauminus}
\tau_- = {\hbar \over 2\pi\alpha \, kT}
{\Omega_0^2 \over \varepsilon_0^2}.
\end{equation}
We stress again that $\tau_-$ determines the diffusion time of the
global phase $\chi$ as it is given in terms of the spectral function
$S_{\chi\chi}(\omega)$.  In the quantum limit $T=0$, the phase diffusion 
time $\tau_-$ diverges.  Furthermore, it follows from
Eq.~(\ref{deph:tauminus}) that $\tau_-$ is {\em tunable}, since
$\varepsilon_0$ may be varied by an external DC-bias.  In particular,
at resonance $\varepsilon_0=0$, the phase diffusion time $\tau_-$
diverges for any temperature.

\begin{figure}[ht]
\centerline{\epsfysize6cm\epsfbox{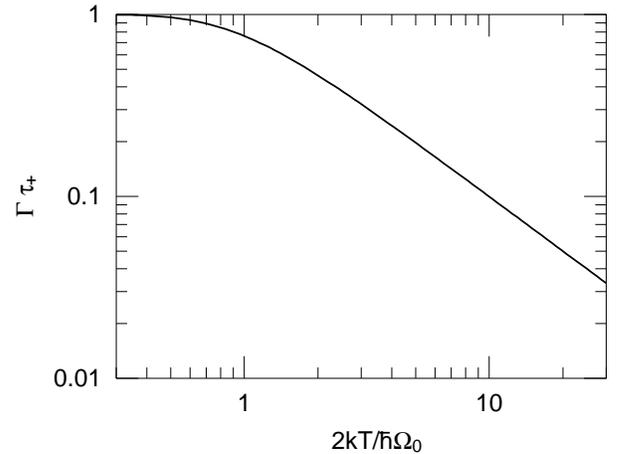}}
\vspace*{1em}
\caption{\label{deph:fig:tplus}The characteristic 
phase diffusion time $\tau_+$ normalized
by the decay rate $\Gamma$, defined in Eq.~(\ref{deph:Gamma}), as a
function of $2kT/\hbar\Omega_0$ in a logarithmic scale.  For high
temperatures, $\tau_+$ is inversely proportional to the temperature
$T$.  For low temperatures, $kT \ll \hbar\Omega_0$, it saturates at a
value $\tau_+=\Gamma^{-1}$.}
\end{figure}

The long time behavior of Eq.~(\ref{deph:cplus}), on the other hand,
is determined by the frequencies near $\Omega_0$.  In the
vicinity of this characteristic frequency, $S_{\varphi\varphi}
(\Omega_0+\omega)$ 
as well as $S_{\theta\theta} (\Omega_0+\omega)$
show a $\omega^{-2}$ behavior {\em at finite as well as at zero
temperature}, which is cut off by the relaxation rate $\Gamma$,
defined in Eq.~(\ref{deph:Gamma}) at very small frequencies $\omega
\sim \Gamma \ll \Omega_0$.  In summary, we have, for $|\omega| \ll
\Omega_0$
\begin{equation}
S_{\varphi\varphi}(\Omega_0+\omega)
\approx
{2\pi\alpha \, \Omega_0 \coth {\hbar\Omega_0 \over 2kT}
\over \omega^2 + (\Gamma/2)^2},
\end{equation}
and $S_{\theta\theta} (\Omega_0+\omega) \approx (\Delta_0^2 /
\Omega_0^2) S_{\varphi\varphi} (\Omega_0+\omega)$.  The time evolution
of Eq.~(\ref{deph:cplus}) for times much larger than the inverse of
the characteristic frequency $\Omega_0$, yet smaller than the inverse
of the relaxation rate $\Gamma$, is therefore linear in time with a
characteristic time $\tau_+$, where
\begin{equation}
\label{deph:tauplus}
\tau_+ = {1 \over \Gamma}
\tanh {\hbar\Omega_0 \over 2kT}.
\end{equation}
Note that Eq.~(\ref{deph:tauplus}) holds for finite temperatures as
well as in the quantum limit.  The phase diffusion time $\tau_+$ is
inversely proportional to $T$ at high temperatures,
\begin{equation}
\label{deph:taupf}
\tau_+ = {1 \over \Gamma} {\hbar\Omega_0 \over 2kT}, \quad
(kT \gg \hbar\Omega_0),
\end{equation}
just as the other characteristic time $\tau_-$.  In the low temperature or
quantum limit, however, it {\em saturates\/} to a value
\begin{equation}
\label{deph:taupq}
\tau_+^{(0)} = {1 \over \Gamma}, \quad (kT \ll \hbar\Omega_0),
\end{equation}
see also Fig.~\ref{deph:fig:tplus}.  The crossover from high
temperature behavior to the quantum limit behavior takes place at $kT
\sim \hbar\Omega_0$.  We point out that $\tau_+$ is related to the
spectral densities $S_{\varphi\varphi}(\omega)$ and
$S_{\theta\theta}(\omega)$ of the internal phases $\varphi$ and
$\theta$.  This indicates that there is a relation between dephasing
(at finite temperature) and the reduction of the persistent current at
zero temperature which is determined by $S_{\varphi\varphi}(\omega)$.

We point out that Eq.~(\ref{deph:cminus}) and Eq.~(\ref{deph:cplus})
do not hold for arbitrarily long times.  In reality the mean square
displacements $\langle |c_\pm(t)-c_\pm(0)|^2 \rangle$ are bounded
since the wave function $\psi_R(t)$ is normalized to 1.  The fact that
$\langle |c_-(t)-c_-(0)|^2 \rangle$, Eq.~(\ref{deph:cminus}), grows
without bounds is an artifact of the linearization of
Eqs.~(\ref{deph:central1})--(\ref{deph:chideq}) and
Eq.~(\ref{deph:central3}).

In summary, we find two characteristic phase diffusion times $\tau_-$
and $\tau_+$, related to the projection of the equilibrium state
$\psi(t)$ onto the ground state and the excited state.  The time
$\tau_-$, associated with the projection on the ground state, is
related to the loss of coherence in the global phase $\chi$ and
diverges as the temperature goes to zero.  The time $\tau_+$, on the
other hand, characterizes the loss of coherence in the internal phases
$\varphi$ and $\theta$, and saturates to a finite value in the limit
of zero temperature. This indicates that even at zero temperature,
coupling to the external circuit causes excitations of the ring into
the excited state. In the zero-temperature limit these excitations
still decay in a finite time.

\section{Conclusion}
We have investigated the persistent current in a normal metal ring
coupled to a resistive external circuit using a Langevin equation
approach.  We have shown that the quantum fluctuations in the external
circuit suppress the persistent current at zero temperature, thus
confirming an earlier Bethe ansatz result.\cite{cedraschi:prl}
Within the same framework, but at finite temperature, we have derived
two dephasing times $\tau_-$ and $\tau_+$, which are the phase
diffusion times of the projections of the wave function of the ring to
ground state and the excited state, respectively, of a ring which is
disconnected from the external circuit.  We show that $\tau_+$ is
related to the spectral densities of the ``internal'' phases $\varphi$
and $\theta$ of the wave function of the ring, which are also
responsible for the reduction of the persistent current.  While
$\tau_-$ diverges in the zero temperature limit, $\tau_+$ saturates to
a finite value.  

\appendix

\section{Coulomb interactions and displacement currents}
\label{coulomb}
In this Appendix, we briefly discuss the electrical coupling of the
ring and the external circuit. The fluctuations of the charge in the
ring and the external circuit are not arbitrary but are connected by
the flow of a displacement current between the ring and the external
circuit. We can express the coupling between the ring and external
circuit in terms of the displacement current. Since the coupling
between these subsystems is crucial, we present first a discussion of
the displacement currents.

In our model the charges and potentials are related via
\begin{eqnarray}
\label{coulomb:Poisson1}
Q_0 &=& C_1 (V_0 - U_d), \\
\label{coulomb:Poisson2}
Q_d &=& C_1 (U_d - V_0) + C_i (U_d - U_a), \\
\label{coulomb:Poisson3}
Q_a &=& C_2 (U_a - V_\infty) + C_i (U_a - U_d), \\
\label{coulomb:Poisson4}
Q_\infty &=& C_2 (V_\infty - U_a),
\end{eqnarray}
where we have introduced the parallel capacitance $C_i \equiv
C_L+C_R$.  In the following, we shall also need the external (serial)
capacitance $C_e^{-1} \equiv C_1^{-1} + C_2^{-1}$, as well as the
parallel and the serial total capacitances $C \equiv C_i + C_e$ and
$C_0^{-1} \equiv C_i^{-1} + C_e^{-1}$.  The whole structure, that is,
the ring together with the external circuit is charge neutral.  As a
matter of fact, it follows from
Eqs.~(\ref{coulomb:Poisson1})--(\ref{coulomb:Poisson4}) that $Q_0 +
Q_d + Q_a + Q_\infty = 0$.  The ring and the external circuit taken
separately, however, do not need to be neutral.  The total charge of
the ring $Q_d+Q_a$ is balanced by the charge on the external
capacitors $C_1$ and $C_2$, $Q_0 + Q_\infty = -(Q_d + Q_a)$, as
follows from Eqs.~(\ref{coulomb:Poisson1})--(\ref{coulomb:Poisson4}).
As the ring and the external circuit do not exchange particles, the
charges on the ring and in the external circuit are conserved, and it
is more convenient to consider the deviations from a reference state.
We denote these deviations by $\delta Q_d$ on the dot and by $\delta
Q_a$ on the arm, and similarly by $\delta Q_0$ and $\delta Q_\infty$
for the external circuit.  They obey the relations $\delta Q_d +
\delta Q_a = \delta Q_0 + \delta Q_\infty=0$.  For time derivatives,
it is of course irrelevant whether we consider total charges or
deviations.

The current flowing out from the capacitor $C_1$ is a pure
displacement current
\begin{equation}
\label{coulomb:I1Q0}
I_1 = -\dot{Q}_0
= C_1 {\partial \over \partial t} \left( U_d-V_0 \right),
\end{equation}
whereas the currents flowing through the right and the left junction,
respectively, are particle currents $I_{L/R}^p$ augmented by
displacement currents
\begin{eqnarray}
\label{coulomb:IL}
I_L &=& I_L^p
+ C_L {\partial \over \partial t} (U_d - U_a), \\
\label{coulomb:IR}
I_R &=& I_R^p
+ C_R {\partial \over \partial t} (U_d - U_a).
\end{eqnarray}
The particle currents are related to the charge on the dot by $I_L^p +
I_R^p = -\dot{Q}_d$.  At each node we have $I_1 + I_L + I_R = I_2 -
(I_L + I_R) = 0$, and thus $I_1 + I_2 = 0$.  These equations
correspond to the law of current conservation at the nodes of an
electrical network.

We can express all electrical quantities of interest in terms of the
external bias $\delta V \equiv V_0-V_\infty$ and the charge deviation,
$\delta Q_d$.  We combine Eqs.~(\ref{coulomb:Poisson2}) and
(\ref{coulomb:Poisson3}) and the condition $\delta Q_d + \delta Q_a =
0$ to obtain the potential difference inside the ring
\begin{equation}
\label{coulomb:UdUa}
U_d - U_a = {C_e \over C} \delta V + {\delta Q_d \over C}.
\end{equation}
The charge deviation $\delta Q_0$ on the capacitor $C_1$ is found from
Eqs.~(\ref{coulomb:Poisson1}), (\ref{coulomb:Poisson4}) and
(\ref{coulomb:UdUa}),
\begin{equation}
\delta Q_0 = C_0 \delta V - {C_0 \over C_i} \delta Q_d,
\end{equation}
where we have used the identity
\begin{equation}
{C_e \over C} = {C_0 \over C_i},
\end{equation}
whence we obtain for the current flowing in the external circuit
\begin{equation}
\label{coulomb:I1}
I_1 = {C_0 \over C_i} \dot{Q}_d
- C_0 \delta\dot{V}.
\end{equation}
In the circuit containing a current noise source, see
Fig.~\ref{deph:noise}, the current $I_1$ flowing out of the circuit
augmented by the current $\delta I$ coming from the noise source must
equal the current through the resistor $I=\delta V/R$, namely
\begin{equation}
\delta V/R = I_1 + \delta I.
\end{equation}
Together with Eq.~(\ref{coulomb:I1}), this gives
Eq.~(\ref{deph:voltagedeq}).

\section{The circulating current}
\label{circ}
The equilibrium persistent current $I$ is a quantum and statistical
average which can be obtained from the derivative of the free energy
$I=-c\,\partial F/\partial\Phi$.  Some discussion is required, if we
are, as in Ref.~\onlinecite{cedraschi:prl} concerned with current
fluctuations.  Naively, we might want to investigate the fluctuations
of the persistent current by considering the second order derivative
of the thermodynamic potential. However, such a procedure works only
if the observable of interest commutes with the Hamiltonian.
Moreover, as pointed out above, the true and physically relevant
currents is the total current (particle current plus displacement
current).  Whereas the displacement current needs not to be considered
as long as we are interested in average quantities only, this is not
true, if we consider fluctuations of the current.  For the model
considered here, we can derive expressions for the current operators
which are particle operators weighted according to the distribution of
the geometrical capacitances of the system.

Consider first for a moment the electrically isolated loop.  In this
case the capacitances to the exterior circuit are $C_{1} = C_{2} = 0$.
From Eq.~(\ref{coulomb:Poisson2}) we find $Q_{d} = C_{i}(U_{d}-U_{a})$
and thus we can also write the currents through the left and right
junctions, Eqs.~(\ref{coulomb:IL}) and (\ref{coulomb:IR}), in terms of
the charge on the dot as, $I_{L} = I_{L}^{p} + (C_{L}/C_{i})
dQ_{d}/dt$ and $I_{R} = I_{R}^{p} + (C_{R}/C_{i}) dQ_{d}/dt$. Using
particle conservation, $I_{L}^{p}+ I_{R}^{p} - dQ_d/dt = 0$, to
eliminate the time-derivative of the charge we find immediately that
the current {\em circulating} in this loop is, $\hat{I}_c \equiv
\hat{I}_L = - \hat{I}_R$ with
\begin{equation}
\label{circ:Ic}
\hat{I}_c = 
{C_R \hat{I}_{L}^{p} - C_L \hat{I}_{R}^{p} \over C_i},
\end{equation}
where $C_i = C_L+C_R$.  The circulating current is thus in general not
determined by the particle currents but by an average of these
currents weighted according to the Coulomb interaction (capacitance
ratios).  Expressions of this type are familiar form the dynamic
transport through double barriers, but seem to be novel for persistent
currents.  We emphasize that Eq.~(\ref{circ:Ic}) does not mean that
the particle currents can now be calculated from the non-interacting
problem.  The dynamic particle currents depend on the self-consistent
potential distribution. For an illustration of this point we refer the
reader to Ref.~\onlinecite{bb} where dynamic current noise spectra for
double barriers are compared based on calculations using the particle
currents of the non-interacting problem and calculations using the
particle currents of the interacting problem.

Next let us consider the system coupled to the external circuit.  In
this case $C_{1} > 0$ and $C_{2} > 0$. We can proceed as above.  We
first express the time derivative of potential difference $U_d - U_a$
in terms of the time derivative of the charge on the dot and the
current $I_{1} = dQ_{0}/dt = C_{1} d(V_{0} - U_d)/dt$ (see
Eq.~(\ref{coulomb:Poisson1})). Using current conservation, $I_{1} =
I_{L} + I_{R}$ and particle conservation $I_{L}^{p}+ I_{R}^{p} -
dQ_d/dt = 0$ we can again eliminate the time derivative of the charge
$Q_d$ and find for the currents through the left and right barrier,
\begin{equation}
\label{circ:IL}
\hat{I}_L = \hat{I}_c + (C_L/C_{i}) I_{1}, 
\end{equation}
\begin{equation}
\label{circ:IR}
\hat{I}_R = - \hat{I}_c + (C_R/C_{i}) I_{1}, 
\end{equation}
with $\hat{I}_c$ as given in Eq.~(\ref{circ:Ic}). 
The current $I_{1}$, which can be induced 
with the help of the external circuit, 
is divided onto the two branches of the ring-dot sample also according 
to a capacitance ratio. 

What is observed in a measurement of the magnetization?  It is
important to note, that the external circuit (see
Fig.~\ref{deph:system}) also forms a current loop and depending on the
geometry of the circuit also contributes to the fluctuation of the
magnetization.  Suppose, the circuit of Fig.~\ref{deph:system} is a
planar structure in the $x$-$y$-plane. Let the ring with the in-line
dot enclose an area $A_r$ and let the external circuit (excluding the
ring) enclose an area $A_e$.  For this circuit the magnetic moment can
be viewed as being generated by a current $I_L$ enclosing the area
$A_1$ and by a current $I_1$ enclosing the area $A_e$. The magnetic
moment is thus $m = (1/c) (I_L A_r - I_1 A_e )$.  Of course the
topology of the current distribution matters: If we consider the
external circuit to be above the ring dot structure instead of below
as shown in Fig. \ref{deph:system}, the magnetization is $m = (1/c) (-
I_R A_r + I_1 A_e )$. In these two circuits the external circuit gives
magnetization contribution into opposite directions. The average
magnetization of the two circuits is just $m = (1/c) I_c A_e$. A third
circuit which permits to investigate the magnetization of the ring and
the external circuit separately is a structure in which the ring lies
in the $x$-$y$-plane and the external circuit say in the
$x$-$z$-plane. Then the magnetization is $m_z = (1/c) I_c A_r$ and
$m_y = (1/c) I_1 A_e$.  $I_c$ appears as the most natural
generalization of the equilibrium persistent current. In
Ref.~\onlinecite{cedraschi:prl} we have investigated the fluctuations
of $I_c$.

All the above expressions are based on linear relations between
currents, charges and potentials. All the above expressions are thus
valid also for operators. We now present a specific expression for the
operator of the circulating current used in
Ref.~\onlinecite{cedraschi:prl}.  Let us consider the particle current
operators $\hat{I}_L^p$ through the left and $\hat{I}_R^p$ through the
right tunnel barrier.  In terms of the Hamilton operators $\hat{H}_L$
and $\hat{H}_R$ responsible for the tunneling across the left and the
right tunnel barrier, respectively, they read $\hat{I}^p_{L/R} =
-i/\hbar [\hat{H}_{L/R}, \hat{Q}_d]$.  In other words, they are equal
to the decrease of the charge on the dot per unit of time through the
right and left junction.  For the two-level system, in a basis in
which the Hamiltonian is real (see Eq.~(\ref{deph:Hring})) we have
\end{multicols}
\widetext
\vspace*{-0.2truein} \noindent \hrulefill \hspace*{3.6truein}
\begin{eqnarray}
\label{circ:ILp}
\hat{I}_L^p &=& -{e \over \hbar}
{\pm t_L t_R \sin (2\pi\Phi/\Phi_0) \, \sigma_x
+ [t_L^2 \pm t_L t_R \cos (2\pi\Phi/\Phi_0)] \sigma_y
\over \sqrt{t_L^2 + t_R^2 \pm 2t_L t_R \cos(2\pi\Phi/\Phi_0)}},\\
\label{circ:IRp}
\hat{I}_R^p &=& {e \over \hbar}
{\pm t_L t_R \sin (2\pi\Phi/\Phi_0) \, \sigma_x
- [t_R^2 \pm t_L t_R \cos (2\pi\Phi/\Phi_0)] \sigma_y
\over \sqrt{t_L^2 + t_R^2 \pm 2t_L t_R \cos(2\pi\Phi/\Phi_0)}}.
\end{eqnarray}
\hspace*{3.6truein} \noindent \hrulefill
\begin{multicols}{2}
\narrowtext
\noindent
Using the particle current operators in Eq.~(\ref{circ:Ic}) leads to
the operator given by Eq.~(\ref{icop}).  With the help of this
operator we can investigate the average persistent current and the
fluctuations (see Eq.~(7) of Ref.~\onlinecite{cedraschi:prl}).


\begin{thebibliography}{10}
\bibitem{note2} Weak coupling gives rise to a small coupling coefficient 
in the Coulomb energy if it is expressed in terms of potentials. If the 
Coulomb energy is written in terms of the polarization charges 
weak coupling corresponds to a divergent coupling term in the Coulomb
energy.
\end{thebibliography}

\end{multicols}

\end{document}